\documentclass[twocolumn,showpacs,preprintnumbers,amsmath,amssymb,eqsecnum]{revtex4}

\usepackage{graphicx}
\usepackage{dcolumn}
\usepackage{bm}

\newcommand{\bra}[1]{\langle{#1}|}
\newcommand{\ket}[1]{|{#1}\rangle}
\newcommand{\erf}[1]{Eq.~(\ref{#1})}

\begin{document}


\title{Approximate Master Equations for Atom Optics}

\author{D.J. Atkins}
 \email{D.Atkins@gu.edu.au}
\author{H.M. Wiseman}%
 \email{H.Wiseman@gu.edu.au}
\author{P. Warszawski}
 \email{prahlad@midway.uchicago.edu}
\affiliation{Centre For Quantum Dynamics, School of Science,
Griffith University, Nathan, Queensland, Australia. 4111}

\date{\today}

\begin{abstract}
In the field of atom optics, the basis of many experiments is a
two level atom coupled to a light field. The evolution of this
system is governed by a master equation. The irreversible
components of this master equation describe the spontaneous
emission of photons from the atom. For many applications, it is
necessary to minimize the effect of this irreversible evolution.
This can be achieved by having a far detuned light field. The
drawback of this regime is that making the detuning very large
makes the time step required to solve the master equation very
small, much smaller than the time scale of any significant
evolution. This makes the problem very numerically intensive. For
this reason, approximations are used to simulate the master
equation which are more numerically tractable to solve. This paper
analyses four approximations: The standard adiabatic
approximation; a more sophisticated adiabatic approximation (not
used before); a secular approximation; and a fully quantum
dressed-state approximation. The advantages and disadvantages of
each are investigated with respect to accuracy, complexity and the
resources required to simulate. In a parameter regime of
particular experimental interest, only the sophisticated adiabatic
and dressed-state approximations agree well with the exact
evolution.
\end{abstract}

\pacs{32.80.Lg, 03.75.Be, 42.50.Vk, 02.60.Cb}
\keywords{Master Equation, Atom Optics, Approximation, Adiabatic,
Fully Quantum, Dressed State, Secular, Large Detuning}
\maketitle

\section{\label{Intro}The Master Equation}

Many investigations in the realm of quantum physics are based on a
single two level atom coupled to an approximately resonant light
field. One major aspect of this work is to investigate the quantum
mechanical motion of such an atom. The work of Steck {\it et al}
\cite{Steck} is a prime example. They investigate chaos assisted
tunneling by studying the motion of cold cesium atoms in an
amplitude modulated standing wave of light. Haycock {\it et al}
\cite{Haycock} study cesium atoms in an effort to observe quantum
coherent dynamics. The work of Hensinger {\it et al} investigates
quantum chaos \cite{Hen01a} and quantum tunneling \cite{Hen01b}.
Proposals in QED to utilize interactions between atoms and their
cavity for physical realizations of a quantum computer
\cite{Pellizzari} and for communication of quantum states
\cite{Cirac} also require good descriptions of quantum mechanical
motion.

The full quantum system here consists of the light field and the
atom. The evolution of such a system is governed by a
Schr\"odinger equation. In most cases however, the evolution of
the light field is not of interest and so the Schr\"odinger
equation can be reduced to a master equation. The most general
form of this master equation is given by
\begin{equation}
\dot{\rho}=-\frac{i}{\hbar}\left[H,\rho\right]+\mathcal{L}
\rho,\label{lindblad}\end{equation} where $\mathcal{L}$ is a
Lindbladian superoperator \cite{Lind}.

The form we will use here, specific to a two level atom
interacting with a light field, is
\begin{subequations}\label{majorME} \begin{equation}
H=\frac{p_x^2}{2m}+\hbar\delta\sigma^\dagger\sigma +
\frac{\hbar}{2}\left[\Omega(x,t)\sigma^\dagger+\Omega^\dagger(x,t)
\sigma\right]\label{Hamiltonian},\end{equation}\begin{equation}
\mathcal{L}\rho=\Gamma\left(\mathcal{B} \mathcal{J}\left[
\sigma\right] \rho -\mathcal{A}\left[\sigma\right]\rho\right)
.\label{Louivillian}\end{equation}\end{subequations} Here we are
only interested in the motion of the atom in the one direction.
Thus $p_x$ is the component of the atom's momentum in the
$x$-direction. The detuning of the light field is $\delta$. The
atomic lowering operator is given by $\sigma=\ket{g}\bra{e}$ where
$\ket{g}$ and $\ket{e}$ represent the ground and excited states of
the atom respectively. $\Omega(x,t)$ is the complex, and possibly
time-dependent Rabi frequency operator for the light field, which,
from here on, will be represented simply as $\Omega$. $k$ and $m$
are the wavenumber of the incident photons and the mass of the
atom respectively. Planck's constant ($\hbar$) will be set to one
for the rest of the analysis. The superoperator $\mathcal{B}$
describes random momentum kicks. It is defined for any arbitrary
operator $a$ as
\begin{equation} \mathcal{B}a= \int_{-1}^1W(u)e^{ikux}ae^{-ikux}du
, \end{equation} where $W(u)$ is the atomic dipole radiation
distribution function produced by the electronic transition,
reduced to one dimension \cite{Hen01a}. For motion parallel to the
direction of propagation of the laser light, it is given by
\begin{equation}
W(u)=\frac{3}{8}\left(1+u^2\right).\label{W(u)}\end{equation} The
superoperators $\mathcal{J}$ and $\mathcal{A}$ are defined for
arbitrary operators $c$ and $a$ as \begin{equation}
\mathcal{J}\left[c\right]a=cac^\dagger,\mathcal{A}\left[c\right] a
=\frac{1}{2}\left\{c^\dagger c,a\right\},\end{equation} and define
the general form of a Lindbladian \cite{Lind} superoperator.

The entire expression of \erf{Louivillian} describes the
irreversible evolution of the system at rate $\Gamma$. It is this
part of the master equation that we wish to minimize by working in
the regime $\delta\gg\Omega_\mathrm{max}\gg\Gamma$ (where $\Omega
_\mathrm{max}$ is the maximum modulus of the Rabi frequency
operator). The problem with this regime is that when making
$\delta$ very large, the full master equation still has to be
solved using a timestep smaller than $\delta^{-1}$. This makes
solving the full master equation numerically very difficult.

There are a number of ways to approximate the master equation,
four of which are investigated here. These are \begin{enumerate}
\item The standard adiabatic approximation; \item A more
sophisticated adiabatic approximation; \item A secular
approximation; and \item A dressed-state approximation.
\end{enumerate} Approximation (1) is a standard approach used by
many researchers in the field, both as a semi-classical treatment
\cite{Parkins} and as a fully quantum approximation method
\cite{Hen01a,Graham}. Unfortunately, this treatment is not valid
in the regime of the work of \cite{Hen01a}. These experiments were
performed in the regime of $\delta\gg\Omega_\mathrm{max}\gg\Gamma$
but where $\Gamma$ is the same order as
$\Omega^2_\mathrm{max}/\delta$. In their work, approximation (1)
was used but on closer examination, this approximation was seen
only to be valid in the regime
$\Gamma\gg\Omega^2_\mathrm{max}/\delta$. We will concentrate on
the difficult $(\Gamma\sim\Omega^2_{\mathrm{max}}/\delta)$ regime
in this analysis. Approximation (2) is one way to correct the
standard adiabatic approximation for this regime. Approximation
(3) was proposed in Ref. \cite{dyrtingmilburn}, and was used for
the formation of quantum trajectory simulations in
Ref.\cite{Hen01a}. Approximation (4) is a fully quantum
dressed-state treatment including the effect of spontaneous
emission. Previously, only semi-classical treatments, both
omitting \cite{Deutschmann}, and including \cite{Dalibard,
Parkins}, spontaneous emission have been done. We look at all four
approximations, examining the validity and complexity, as well as
the numerical accuracy (compared to the full simulation) and
computational resource requirements, of each.

\section{The standard adiabatic approximation\label{standard}}

The method of adiabatic approximation applied to the full master
equation serves to eliminate the internal state structure of the
atom. One reason for wishing to have no internal states is so that
the system has a clear classical analogue \cite{dyrtingmilburn}.
In performing this treatment, we also remove the Hamiltonian term
of order $\delta$, removing the requirement that the master
equation be solved on a timestep at least as small as
$\delta^{-1}$. This adiabatic elimination technique is described
in many different text books, see \cite{Meystre, Barnett}. The
result we obtain was first derived by Graham, Schlautmann and
Zoller \cite{Graham}.
 To achieve this
adiabatic elimination however, we follow a procedure similar to
that of Hensinger {\it et al} \cite{Hen01a}.

The density matrix can be written using the internal state basis
as
\begin{eqnarray} \rho&=&\rho_{gg}\otimes\ket{g}\bra{g} +
\rho_{ge}\otimes\ket{g}\bra{e} \nonumber \\ &
&\rho^\dagger_{ge}\otimes\ket{e}\bra{g}+\rho_{ee}\otimes
\ket{e}\bra{e}, \end{eqnarray} where the $\rho_{gg}$ etc are still
operators on the centre-of-mass Hilbert space
L$^{(2)}(\mathbb{R})$. From Eqs. (\ref{majorME}), these obey
\begin{subequations}\label{rates} \begin{eqnarray}
\dot{\rho}_{gg}&=&
\Gamma\mathcal{B}\rho_{ee}-\frac{i}{2}\left(\Omega^\dagger
\rho_{eg}-\rho_{ge}\Omega\right) -\mathcal{K}\rho_{gg}
,\label{fullrhogg} \\
\dot{\rho}_{ge}&=&-\frac{\Gamma}{2}\rho_{ge}-\frac{i}{2}\left(
\Omega^\dagger\rho_{ee}-\rho_{gg}\Omega^\dagger\right)
+i\delta\rho_{ge}\nonumber
\\ & &-\mathcal{K}\rho_{ge},\label{fullrhoge}
\\ \dot{\rho}_{ee}&=& -\Gamma\rho_{ee}-\frac{i}
{2}\left(\Omega\rho_{ge}-\rho_{eg}\Omega^\dagger\right)
-\mathcal{K}\rho_{ee}.\label{fullrhoee}
\end{eqnarray}\end{subequations} Here the kinetic energy term is
represented in the superoperator \begin{equation}
\mathcal{K}\rho=i\left[
\frac{p_x^2}{2m},\rho\right].\end{equation}

If the standard adiabatic elimination procedure is valid, we can
represent the system by the density matrix for the centre of mass
(com) alone,
\begin{equation}\rho_{\rm com}={\rm
Tr_{int}}[\rho]=\rho_{gg}+\rho_{ee},\end{equation} where ${\rm
Tr_{int}}$ is the trace over the internal states of the atom. In
this standard adiabatic approximation we further simplify this by
noting that large detuning leads to very small excited state
populations such that $\rho_{gg}\gg\rho_{ee}$ and thus $\rho_{\rm
com}$ is approximately $\rho_{gg}$. Thus denoting $\rho_{\rm com}$
simply as $\rho$, we can replace $\rho_{gg}$ in \erf{fullrhogg}
just with $\rho$.

Now we require expressions for $\rho_{ge}$ and $\rho_{ee}$ in
terms of $\rho$. This is achieved by noting that from
\erf{fullrhoge}, $\rho_{ge}$ comes to equilibrium on a timescale
much shorter than $\rho_{gg}$, at a rate $\Gamma/2$. Thus we set
$\dot{\rho}_{ge}= 0$. Also, if the kinetic energy term is much
smaller than $\delta$ then it can be ignored. This will be the
case if $\delta\gg\langle p_x^2/m\rangle$. This is typically true,
and so we get
\begin{equation} \rho_{ge}=i\frac{\rho_{gg}\Omega^\dagger-
\Omega^\dagger\rho_{ee}}{\Gamma-2i\delta}.\label{halfrhoge}
\end{equation}

This can be substituted back into the equation for $\rho_{ee}$ to
give \begin{eqnarray} \dot{\rho}_{ee}&=&-\Gamma\left(\rho_{ee}
+\frac{\left\{\Omega\Omega^\dagger,\rho_{ee}\right\}} {2\left(
\Gamma^2+4\delta^2\right)} +\frac{i\delta\left[\Omega\Omega^
\dagger ,\rho_{ee}\right]}{\Gamma\left(\Gamma^2+4\delta^2 \right)}
\right)\nonumber \\ & &
+\frac{\Gamma\mathcal{J}\left[\Omega\right]\rho_{gg}}{\Gamma^2
+4\delta^2} -\mathcal{K}\rho_{ee}.\label{laterhoee}\end{eqnarray}

From \erf{laterhoee}, we see that $\rho_{ee}$ equilibrates on a
timescale much faster than $\rho_{gg}$ and so we also set
$\dot{\rho}_{ee}= 0$. This time, the kinetic energy term must be
ignored compared with $\Gamma$ rather than $\delta$. Allowing this
approximation gives
\begin{eqnarray}\rho_{ee}+\frac{\left\{\Omega\Omega^
\dagger,\rho_{ee}\right\}}{2\left(\Gamma^2+
4\delta^2\right)}+\frac{i\delta
\left[\Omega\Omega^\dagger,\rho_{ee}\right]}{\Gamma\left(\Gamma^2+
4\delta^2\right)} \simeq\frac{\Omega\rho_{gg}\Omega^\dagger}
{\Gamma^2+4\delta^2}\label{badrhoee}.\end{eqnarray}The first
correction term on the left hand side scales as
$(\Omega/\delta)^2$ which in the regime chosen is negligible
compared to the leading order term. The second correction term
however scales as $\left(\Omega^2/\delta\right)/\Gamma$. Had we
been working in the regime $\Gamma\gg\Omega^2/\delta$, then this
term could also be safely ignored compared to the leading term.
This condition however, is not satisfied in the experiments of
\cite{Hen01a} nor in our chosen regime, leaving this term the same
order as the leading term. This term however was dropped in Ref.
\cite{Hen01a} on the basis that in a more sophisticated approach
(Sec. \ref{better}), this term does not appear and the correction
to the final master equation is small \cite{Hen01a}. Knowing that
this is an unjustified approximation, but in the interest of
comparison to currently used techniques \cite{Hen01a}, we will
continue to follow this method as others have done. Thus, dropping
the second correction term we are left with
\begin{equation} \rho_{ee}\simeq\frac{\Omega
\rho_{gg}\Omega^\dagger} {\Gamma^2+4\delta^2}.
\label{finalrhoee}\end{equation}

Now substituting first \erf{halfrhoge} then \erf{finalrhoee} into
\erf{fullrhogg} and replacing $\rho_{gg}$ with $\rho$ gives the
final adiabatically eliminated master equation:\begin{eqnarray}
\dot{\rho}&=&\Gamma\left(\mathcal{B} \mathcal{J}\left[
\frac{\Omega}{2\delta}\right]\rho
-\mathcal{A}\left[\frac{\Omega}{2\delta}
\right]\rho\right)-i\left[\frac{p_x^2}{2m}-\frac{\Omega\Omega
^\dagger}{4\delta},\rho\right],\nonumber \\
\label{adiab1me}\end{eqnarray} where here we have used the fact
that $\delta\gg\Omega\gg\Gamma$ to eliminate some of the higher
order terms. This master equation is of the Lindblad form.

\section{A more sophisticated Adiabatic approximation\label{better}}

As noted, the standard adiabatic elimination method described in
section \ref{standard} is not strictly valid in the regime of the
experiments of Hensinger {\it et al} \cite{Hen01a},
$\Gamma\sim\Omega_\mathrm{max}^2/\delta$. There are a number of
ways to try and develop a strictly valid version of the adiabatic
approximation in this regime. One way would be to not drop any
terms without justification and continue to plough through the
mathematics. Another method which we believe to be neater and just
as accurate is proposed here by using a slightly more
sophisticated method similar to that in the appendix of
\cite{howpolly}.

The basis of the approach is to move into an interaction picture
with respect to\begin{equation}H_0=\frac{\Omega\Omega
^\dagger}{4\delta} \label{H0}.\end{equation} This approach may
seem counter-intuitive to most. Usually when moving to an
interaction picture, it would be with respect to a $H_0$ that is
already one of the terms in the Hamiltonian. In our case, if we
investigate \erf{adiab1me}, we are moving to an interaction
picture with respect to the opposite of a term in the effective
Hamiltonian and as such will actually be adding a term to the
Hamiltonian. The reason we choose to do this is that the problem
term we encountered in Sec. \ref{standard} was in the Hamiltonian
for the excited state. The potential seen by the excited state of
the atom is in fact inverted and so the chosen $H_0$ is designed
to cancel the excited state potential.

After moving into this interaction picture, we then perform the
adiabatic elimination process, and finally transform back into the
Schr\"odinger picture. This method can give a different result
because the approximations we make in the interaction picture may
not have been valid in the Schr\"odinger picture.

With the unitary transformation operator $U_0(t)=e^{-iH_0t}$, the
interaction picture density matrix is
\begin{equation}\tilde{\rho}=U^\dagger_0(t)\rho U_0(t)
\label{IPrho}. \end{equation} Using this in Eqs. (\ref{majorME})
gives an interaction picture master equation equation still of the
Lindblad form
\begin{equation} \dot{\tilde{\rho}}=-i\left[\tilde{V}
,\tilde{\rho}\right]+\Gamma\mathcal{L}\tilde{\rho},\end{equation}
but now with an extra Hamiltonian term such that $\tilde{V}$ is
given by \begin{eqnarray}
\tilde{V}=\frac{\tilde{p}_x^2}{2m}&+&\delta\sigma^\dagger \sigma +
\frac{1}{2}\left(\Omega\sigma^\dagger+\Omega^\dagger
\sigma\right)+\frac{\Omega\Omega ^\dagger}{4\delta}
\label{VIP}.\end{eqnarray}Where $\tilde{p}_x$ is just the
component of the momentum in the $x$-direction, transformed into
the interaction picture. The Lindbladian superoperator
$\mathcal{L}$ is unaffected by the interaction picture because
Rabi frequency operator $\Omega$ commutes with the position
operator, $x$, as well as with the state operators, $\sigma$ and
$\sigma^\dagger$.

Following the same procedure as the standard adiabatic elimination
process, the equations for the centre-of-mass operators can be
extracted:
\begin{subequations}\label{IPREs}\begin{eqnarray}
\dot{\tilde{\rho}}_{gg}&=&
\Gamma\mathcal{B}\tilde{\rho}_{ee}-\frac{i}{2}\left(\Omega^\dagger
\tilde{\rho}_{eg}-\tilde{\rho}_{ge}\Omega\right)
-\frac{i}{4\delta}\left[\Omega\Omega^\dagger,
\tilde{\rho}_{gg}\right]\nonumber \\
& &-\tilde{\mathcal{K}}\tilde{\rho}_{gg}, \label{fullrhogg2}\\
\dot{\tilde{\rho}}
_{ge}&=&-\frac{\Gamma}{2}\tilde{\rho}_{ge}-\frac{i}{2}\left(
\Omega^\dagger\tilde{\rho}_{ee}
-\tilde{\rho}_{gg}\Omega^\dagger\right)
-\frac{i}{4\delta}\left[\Omega\Omega^\dagger,\tilde{\rho}_{ge}
\right]\nonumber \\ & &
+i\delta\tilde{\rho}_{ge}-\tilde{\mathcal{K}}
\tilde{\rho}_{ge} ,\label{fullrhoge2} \\
\dot{\tilde{\rho}}_{ee}&=& -\Gamma\tilde{\rho}_{ee}-\frac{i}
{2}\left(\Omega\tilde{\rho}_{ge}-\tilde{\rho}_{eg}
\Omega^\dagger\right)
-\frac{i}{4\delta}\left[\Omega\Omega^\dagger,
\tilde{\rho}_{ee}\right]\nonumber \\ &
&-\tilde{\mathcal{K}}\tilde{\rho}_{ee} ,\label{fullrhoee2}
\end{eqnarray}\end{subequations} where obviously $\tilde{
\mathcal{K}}$ uses the interaction picture momentum operator
$\tilde{p}_x$.

In this more sophisticated approach, we still take the trace over
the internal states of the atom, letting $\tilde{\rho}_{\rm
com}=\tilde{\rho}_{gg}+\tilde{\rho}_{ee}$, but now we do not
simplify this further and simply let
$\tilde{\rho}=\tilde{\rho}_{gg}+\tilde{\rho}_{ee}$. This gives a
master equation of the form \begin{eqnarray}
\dot{\tilde{\rho}}&=&\Gamma\left(\mathcal{B}-1\right)\tilde{\rho}
_{ee} -\frac{i}{2}\left[\Omega^\dagger,\tilde{\rho}_{eg} \right]
-\frac{i}{2}\left[\Omega,\tilde{\rho}_{ge}\right]\nonumber
\\ & &-
\frac{i}{4\delta}\left[\Omega\Omega^\dagger,\tilde{ \rho} \right]
-\tilde{\mathcal{K}}\tilde{\rho}. \label{MEees} \end{eqnarray}
Hence we again need to find expressions for $\tilde{\rho}_{ee}$
and $\tilde{\rho}_{ge}$ in terms of $\tilde{\rho}$. We can achieve
this by noting that, as in the earlier treatment,
$\tilde{\rho}_{ee}$ and $\tilde{\rho}_{ge}$ equilibrate on a
timescale much faster than $\tilde{\rho}_{gg}$. By setting
$\dot{\tilde{\rho}}_{ge}=0$, and again ignoring the kinetic energy
term, we get
\begin{eqnarray} \tilde{\rho}_{ge}&=& -\frac{\tilde{\rho}
\Omega^\dagger}{2\delta}-\frac{\left[\Omega\Omega^\dagger,
\tilde{\rho}\Omega^\dagger\right]}{8\delta^3}  +\frac{
\left\{\Omega^\dagger,\tilde{\rho}_{ee}\right\}}{2\delta}
-\frac{\Gamma\tilde{\rho}\Omega ^\dagger}{4i\delta^2}\nonumber
\\ & &+\frac{\Gamma\left\{\Omega^\dagger,\tilde{\rho} _{ee}
\right\}}{4i\delta^2} +\frac{\left[\Omega\Omega
^\dagger,\left\{\Omega^\dagger,\tilde{
\rho}_{ee}\right\}\right]}{8\delta^3}.\label{IPrhoge5}
\end{eqnarray}

Setting $\dot{\tilde{\rho}}_{ee}=0$ and ignoring the kinetic
energy term allows us to solve for $\tilde{\rho}_{ee}$. In the
more sophisticated adiabatic treatment, instead of getting an
equation of the form of \erf{badrhoee}, we get \begin{equation}
\mathcal{R}\tilde{\rho}_{ee}=\mathcal{F}\frac{\Omega
\tilde{\rho}\Omega^\dagger}{4\delta^2}\label{IPhalfrhoee}.
\end{equation} In \erf{IPhalfrhoee}, the superoperators
$\mathcal{F}$ and $\mathcal{R}$ are defined \begin{eqnarray}
\mathcal{F}&\equiv&1+\frac{i}{2\delta\Gamma}\left[\Omega\Omega
^\dagger,\ldots\right],\label{F}\\\mathcal{R}&\equiv&\mathcal{F}+\text{
higher order terms}.\label{R}\end{eqnarray}

In the regime chosen ($\Gamma\sim\Omega^2/\delta$), the second
term of \erf{F} is of order 1 and so cannot be ignored compared to
the leading term. In \erf{R}, the higher order terms are of order
much smaller than 1 and even smaller than $\Omega^4/\delta^3$,
which is the smallest order kept by making Taylor approximations
to get the expression for $\tilde{\rho}_{ge}$. Thus to simplify
\erf{IPhalfrhoee}, we act on the left of both sides with the
superoperator $\mathcal{F}^{-1}$. Then to leading order we have
\begin{eqnarray} \tilde{\rho}_{ee}&\simeq&\frac{\Omega
\tilde{\rho}\Omega^\dagger}{4\delta^2}. \label{IPrhoee}
\end{eqnarray} As expected, the interaction picture chosen
produced a term in $\rho_{gg}$ to counteract the unwanted term in
$\rho_{ee}$. Notice here that we come to essentially the same
result as in the standard adiabatic treatment in \erf{finalrhoee}
but without making any unjustified approximations.

Now substituting \erf{IPrhoge5} and \erf{IPrhoee} into
\erf{MEees}, after simplification, leads to the final interaction
picture master equation \begin{eqnarray} \dot{\tilde{\rho}} &=&
\Gamma\left(\mathcal{B} \mathcal{J}\left[
\frac{\Omega}{2\delta}\right]\tilde{\rho}
-\mathcal{A}\left[\frac{\Omega}{2\delta}
\right]\tilde{\rho}\right)
-i\left[\frac{\Omega^2{\Omega^\dagger}^2}{16\delta
^3},\tilde{\rho}\right]\nonumber
\\ & &- \tilde{\mathcal{K}}\tilde{\rho}.\end{eqnarray} Finally,
all that remains is to transform the interaction picture master
equation back to the Schr\"odinger picture by performing the
opposite unitary transformation. This leaves the final master
equation for the more sophisticated adiabatic elimination
treatment as\begin{eqnarray}\dot{\rho}&=&\Gamma\left(\mathcal{B}
\mathcal{J}\left[ \frac{\Omega}{2\delta}\right]\rho
-\mathcal{A}\left[\frac{\Omega}{2\delta}
\right]\rho\right)\nonumber \\ & &
-i\left[\frac{p_x^2}{2m}-\frac{\Omega\Omega
^\dagger}{4\delta}+\frac{\Omega^2 {\Omega^\dagger}^2}{16\delta
^3},\rho\right].\label{adiab2me}\end{eqnarray} Notice here that by
explicitly accounting for the term in $\Omega^2/\Gamma\delta$, the
master equation derived is the same as that of the standard
adiabatic approach with an extra potential term of order
$\Omega^4/\delta^3$. This term is very small and as such the
statements made to justify the standard approach \cite{Hen01a}
were correct in that the adjustment to the final master equation
is small. The more sophisticated treatment however, although more
algebraically intensive to produce the initial master equation,
requires very little extra effort than the standard adiabatic
treatment to simulate. More importantly though, the master
equation derived in the more sophisticated adiabatic approach is
valid in the regime $\Omega_\mathrm{max}^2/\delta\sim\Gamma$.

\section{A Secular Approximation \label{secular}}

A secular approximation to the full master equation is quite
different from any adiabatic approximation. A secular
approximation does not totally remove any dependence on the
internal state of the atom. It only eliminates the coherences
between the internal atomic states. The secular approximation was
also used by Hensinger {\it et al} \cite{Hen01a}, alongside the
standard adiabatic approximation.

There are a number of ways to derive a secular approximation to
the full master equation. One such method has been performed by
Dyrting and Milburn \cite{dyrtingmilburn} but this method is quite
complicated. A much simpler method which produces the same
approximate master equation is based on the technique that is
applied for the standard adiabatic approximation.

We start with the Eqs.~(\ref{rates}) for $\rho_{gg}$ etc, and then
adiabatically eliminate $\rho_{eg}$ as in Sec. \ref{standard}.
However, instead of also trying to solve for $\rho_{ee}$, we just
substitute the expression for $\rho_{ge}$, \erf{halfrhoge}, into
the equations for $\rho_{gg}$, \erf{fullrhogg} and $\rho_{ee}$,
\erf{fullrhoee}. This eliminates the coherences between the
excited and ground states while still keeping much of the original
master equation. One advantage of this is that it allows
comparison to an analogous 2-state classical model
\cite{dyrtingmilburn}. More importantly to us, this approximation
eliminates the evolution at rate $\delta$ as is necessary to
simplify the simulations. We are left with the following equations
for the ground and excited state density matrices
\begin{eqnarray} \dot{\rho}_{gg}&=&\Gamma
\mathcal{B}\rho_{ee}  +\Gamma
\left(\mathcal{J}\left[\frac{\Omega^\dagger}{2\delta}\right]\rho_{ee}
-\mathcal{A}\left[\frac{\Omega}{2\delta}\right]\rho_{gg}\right)
\nonumber \\& &+i\left[\frac{\Omega\Omega^\dagger}{4\delta},
\rho_{gg}\right]-\mathcal{K}\rho_{gg} ,\label{secgg}\\
\dot{\rho}_{ee} &=&
-\Gamma\rho_{ee}+\Gamma\left(\mathcal{J}\left[\frac{\Omega}{2\delta}\right]\rho_{gg}
-\mathcal{A}\left[\frac{\Omega^\dagger}{2\delta}\right]\rho_{ee}\right)
\nonumber \\ & &-i\left[\frac{
\Omega\Omega^\dagger}{4\delta},\rho_{ee}\right]-\mathcal{K}\rho_{ee}.
\label{secee}
\end{eqnarray}

The final master equation is constructed by recombining the
equations (\ref{secgg}) and (\ref{secee}) to produce an equation
which reproduces these equations for $\dot{\rho}_{gg}$ and
$\dot{\rho}_{ee}$ while also only giving rapidly decaying terms
for $\dot{\rho}_{ge}$ and $\dot{\rho}_{eg}$. This final master
equation for the secular approximation is
\begin{eqnarray}\dot{\rho}&=&-i\left[\frac{p_x^2}{2m}+
\frac{\Omega\Omega^\dagger}{4\delta}\sigma_z,\rho\right]
+\Gamma\left(\mathcal{B}\mathcal{J}
\left[\sigma\right]\rho-\mathcal{A}\left[\sigma\right]\rho\right)
\nonumber \\ & &+
\Gamma\left(\mathcal{J}\left[\frac{\Omega^\dagger\sigma}{2\delta}
\right]\rho -\mathcal{A}\left[\frac{\Omega^\dagger\sigma}{2\delta}
\right]\rho\right)\nonumber\\& &+\Gamma\left(
\mathcal{J}\left[\frac{\Omega\sigma^\dagger}{2\delta}\right]\rho-
\mathcal{A}\left[\frac{\Omega\sigma^\dagger}{2\delta}\right]\rho
\right) ,\label{secularME}\end{eqnarray} where $\sigma_z$ is just
the Pauli spin operator $\sigma^\dagger\sigma- \sigma\sigma
^\dagger$.

To compare this master equation with those we have already seen,
the Hamiltonian terms derived here are the same as those derived
in the standard adiabatic approximation, \erf{adiab1me}. The
spontaneous emission term exactly as in the full master equation,
\erf{Louivillian}, remains, while two extra jump terms involving a
state change with no spontaneous emission have been derived.

There are no apparent problems in this derivation, or that in
\cite{dyrtingmilburn}. Nevertheless, as we will discuss in Sec
\ref{Discussion}, \erf{secularME} does not give accurate results
in comparison with \erf{lindblad}, in the regime
$\Omega^2\sim\Gamma\delta$.

\section{A Dressed-State Approximation \label{dress}}

A semi-classical dressed-state treatment of atomic motion was put
forward by Dalibard and Cohen-Tannoudji \cite{Dalibard}. The
dressed-state approximation used here is a fully quantum version
of that treatment.

The states we have been using for a basis so far, $\ket{g}$ and
$\ket{e}$, are called bare states. We can also work with another
basis of position-dependent states which we call dressed-states.
These dressed-states are derived by considering the Hamiltonian of
the full master equation \erf{Hamiltonian} and ignoring the
kinetic energy component to get
\begin{equation} V=\delta\sigma^\dagger\sigma +
\mbox{$\frac{1}{2}$}\left(\Omega\sigma^\dagger+\Omega^\dagger
\sigma\right).\end{equation}The diagonalization of the Hamiltonian
yields the form
\begin{equation} V = E_1(x,t)a^\dagger a + E_2(x,t)aa^\dagger,
\end{equation} where $E_1$ and $E_2$ are the eigenenergies of $V$.
The corresponding eigenstates are the position dependent
dressed-states we will be considering here, labeled $\ket{1}$ and
$\ket{2}$. In terms of these dressed-states, $a$ is given by
$\ket{2}\bra{1}$. The energies and states are
\begin{eqnarray} E_1(x,t)&=&\mbox{$\frac{1}{2}$}
\left(\delta+\Delta\right), \\ E_2(x,t)&=&\mbox{$\frac{1}{2}$}
\left(\delta-\Delta\right), \\
\ket{1}&=&\sin{\theta}\ket{g}+\cos{\theta}\ket{e} ,\\
\ket{2}&=&\cos{\theta}\ket{g}-\sin{\theta}\ket{e},
\end{eqnarray} where $\theta$ is defined by \begin{subequations}
\label{theta}\begin{eqnarray}
\sin{\theta}&=&\frac{\Omega}{\sqrt{2}}\left(\delta^2+\delta\Delta
+\Omega\Omega^\dagger\right)^{-\frac{1}{2}} \label{sin},\\
\cos{\theta}&=& \frac{\Delta+\delta}{\sqrt{2}}\left(\delta^2
+\delta \Delta +\Omega\Omega^\dagger\right)^{-\frac{1}{2}}
,\label{cos}\end{eqnarray}\end{subequations} and $ \Delta=
\left(\delta^2+\Omega\Omega^\dagger\right)^{\frac{1}{2}}.$

Now, we use this basis to get equations for $\rho_{11}$,
$\rho_{12}$, $\rho_{21}$ and $\rho_{22}$. Without including the
spontaneous emission or the kinetic energy, these equations are
simply
\begin{equation}\dot{\rho}_{jk}=-i\left[E_j(x,t)\rho_{jk}
-\rho_{jk}E_k(x,t)\right].\label{rhojk}\end{equation}We do however
want to include the spontaneous emission in our treatment and so
we must assess the effect of the raising and lowering operators
for the bare states ($\sigma=\ket{g}\bra{e}$ and
$\sigma^\dagger=\ket{e}\bra{g}$) acting on the new basis states
($\ket{1}$ and $\ket{2}$). This is easily done.

So far there have been no approximations made. The essence of the
dressed-state approximation is similar to the secular
approximation in that we keep the internal dressed-states but
allow the coherences to go to zero. In contrast to the secular
approximation, we do not simply set $\dot{\rho}_{12}=0$. Instead,
we notice from \erf{rhojk} that, if we ignore the operator nature
of the eigenenergies, then the equations for $\rho_{12}$ would
have a term of the form $-i\left(E_1-E_2\right)\rho_{12}$. To
first order, $E_1-E_2$ is approximately $\delta$. Thus $\rho_{12}$
will rotate very quickly such that only terms rotating at this
very rapid pace will be able to contribute to its evolution. Thus
only terms involving $\rho_{12}$ are kept in the equation for
$\rho_{12}$
\begin{eqnarray} \dot{\rho}_{12}&=& -\Gamma
\mathcal{B}\left(\sin\theta\cos\theta\rho_{12}\cos\theta\sin
\theta\right)\nonumber \\ & &
-\frac{\Gamma}{2}\left(\cos^2\theta\rho_{12}+
\rho_{12}\sin^2\theta\right)\nonumber \\ & &
-i\left[E_1(x,t)\rho_{12}-\rho_{12}
E_2(x,t)\right].\label{dress12} \end{eqnarray} Knowing that the
last term in \erf{dress12} serves to force $\rho_{12}$ to
oscillate very rapidly and the other two terms force it to decay
quickly, $\rho_{12}$ will average to zero. Thus we set $\rho_{12}$
and $\rho_{21}$ equal to zero in the population equations giving
\begin{eqnarray} \dot{\rho}_{11} &=&\Gamma\mathcal{B}\left(
\sin\theta\cos\theta\rho_{11}\sin\theta\cos\theta+\sin^2\theta\rho
_{22}\sin^2\theta \right)\nonumber \\ & &
-\frac{\Gamma}{2}\left\{\cos^2\theta,\rho_{11}\right\}
-i\left[E_1(x,t),\rho_{11}\right],\label{dress111}\\
\dot{\rho}_{22} &=&\Gamma\mathcal{B}\left(
\sin\theta\cos\theta\rho_{22}\sin\theta\cos\theta+\cos^2\theta\rho
_{11}\cos^2\theta \right)\nonumber \\ & &
-\frac{\Gamma}{2}\left\{\sin^2\theta,\rho_{22}\right\}
-i\left[E_2(x,t),\rho_{22}\right].\label{dress222}\end{eqnarray}
If we could ignore the operator nature of the rates in Eqs.
(\ref{dress12}-\ref{dress222}), we would obtain the same rates as
given by Dalibard and Cohen-Tannoudji in \cite{Dalibard}.

All that remains now is to design the master equation that forces
$\rho_{12}$ to zero giving these population equations. The master
equation which fulfills these requirements is
\begin{eqnarray} \dot{\rho}&=& -\mathcal{K}\rho-i\left[E_1(x,t) \ket{1}\bra{1}
+E_2(x,t)\ket{2}\bra{2},\rho\right] \nonumber \\ & &
+\Gamma\mathcal{B}\mathcal{J}
\left[\cos\theta\sin\theta\left(a^\dagger a-aa^\dagger
\right)\right]\rho \nonumber \\ & &
-\Gamma\mathcal{A}\left[\cos\theta\sin\theta\left(a^\dagger
a-aa^\dagger \right)\right]\rho\nonumber \\ & &
+\Gamma\left(\mathcal{B}\mathcal{J}\left[\cos^2\theta a
\right]\rho-\mathcal{A}\left[\cos^2\theta a
\right]\rho\right)\nonumber \\ &
&+\Gamma\left(\mathcal{B}\mathcal{J}\left[\sin^2\theta a^\dagger
\right]\rho-\mathcal{A}\left[\sin^2\theta a^\dagger
\right]\rho\right),\end{eqnarray} where the kinetic energy term
has been restored. This form of the master equation is still very
complicated remembering the definitions of $\cos\theta$ and
$\sin\theta$ from Eqs. (\ref{theta}). It is possible to simulate
this master equation as it stands but the simulation would be very
slow and inefficient. This master equation can, however, be
approximated further with almost no loss in accuracy.

The definitions in Eqs. (\ref{theta}) can be approximated by
remembering that we are working in the regime of
$\delta\gg\Omega$. Thus, to leading order, $\cos\theta$ is simply
1, and $\sin\theta$ is $\Omega/2\delta$. Also any terms of order
$\sin^4\theta$ are extremely small and so are also ignored. If we
propagate these approximations through our system, we find that
the dressed-state $\ket{1}$ is very close to the bare excited
state $\ket{e}$. Also, the dressed-state $\ket{2}$ is very near
the bare ground state $\ket{g}$. Thus if we make the
approximations
\begin{equation} \ket{1}\simeq \ket{e}\text{
, }\ket{2}\simeq \ket{g},
\end{equation} then we can similarly approximate
\begin{equation} \sigma\simeq a,
\sigma^\dagger\simeq a^\dagger .\end{equation}

The last approximation is to expand the dressed-state
eigenenergies in a Taylor series to second order giving
\begin{subequations}\begin{eqnarray}E_1(x,t)&\simeq& \delta+
\frac{\Omega\Omega^\dagger}{4\delta}
-\frac{\Omega^2{\Omega^\dagger}^2}{16\delta^3},\\
E_2(x,t) &\simeq& -\left(\frac{\Omega\Omega^\dagger}{4\delta}
-\frac{\Omega^2{\Omega^\dagger}^2}{16\delta^3}\right).
\end{eqnarray}\end{subequations}

This leaves us with the final master equation for the
dressed-state approximation as \begin{eqnarray} \dot{\rho}&=&
-i\left[\frac{p_x^2}{2m}+\left(\frac{\Omega\Omega^\dagger}
{4\delta}-\frac{\Omega^2{\Omega^\dagger}^2}{16\delta^3}\right)
\sigma_z,\rho\right] \nonumber
\\ & & +\Gamma\left(\mathcal{B}\mathcal{J}
\left[\sigma\right]\rho-\mathcal{A}\left[\sigma\right]\rho\right)
\nonumber \\ & & +\Gamma\left(\mathcal{B}\mathcal{J}
\left[\frac{\Omega}{2\delta}\sigma_z\right]\rho
-\mathcal{A}\left[\frac{\Omega}{2\delta}\sigma_z\right]\rho\right).
\label{finaldressME}\end{eqnarray} Here we have removed the
$H_0=\delta\sigma^\dagger\sigma$ Hamiltonian term by moving into
an interaction picture.

Again comparing this master equation to those already seen, the
Hamiltonian terms here are exactly analogous to those derived in
the sophisticated adiabatic approximation \erf{adiab2me}. Again we
have the same spontaneous emission term as in \erf{Louivillian},
but this time, the higher order correction term includes a
spontaneous emission without changing the internal state of the
atom, the opposite of the case in the secular approximation.

This would be a useful approximate master equation, especially if
the atom was initially in the excited state. In our case however,
we can simplify this equation further by noting that the jump
terms keep the atom in the ground state. Once the excited state
populations are reduced to zero, this master equation reduces to
exactly the same master equation as derived for the more
sophisticated adiabatic approximation \erf{adiab2me}. The dressed
state master equation results would thus lie exactly on top of
those of the more sophisticated adiabatic approximation and as
such are not included in the simulations.

\section{Simulation Of The Master Equations\label{Simulation}}

Now that we have derived the forms of the master equations which
we wish to compare, we need to set up a method of simulation for
the different approximations and the full master equation. In this
case, the numerical environment MATLAB turned out to be the most
useful tool, combined with the Quantum Optics Toolbox produced by
S.M. Tan \cite{Sze1,Sze2}. The simulation is designed by
converting states and operators to vectors and matrices. Making
this conversion requires a number of different adaptations of the
theoretical master equations.

Firstly, we need to chose a form for the complex Rabi frequency
operator, $\Omega$. The form of the Rabi frequency operator we use
here is $\Omega=\Omega_\mathrm{max}\sin kx$, such that there is no
time dependence (standing wave) and the operator is Hermitian.
Rewriting the sine function in terms of exponentials, $e^{\pm
ikx}$, we can examine the action of these exponentials. This
suggests using the momentum representation, because the
exponentials, $e^{\pm ikx}$, simply represent single momentum
kicks to the atom of $\mp1 \hbar k$, or $\mp1$ atomic unit of
momentum.

Having chosen to use the momentum representation to evaluate our
master equation solutions, we need to address concerns with the
conversion process. Firstly, the momentum range must be truncated.
The loss of probability due to this truncation should be kept
below some small level, say $10^{-5}$. Using this rule of thumb,
it was found that a momentum Hilbert space ranging from $-25\hbar
k$ up to $+25\hbar k$ was sufficient for the parameters we chose
(discussed later).

We still face a limitation problem in that the momentum Hilbert
space is continuous. To simulate this on a computer, we need to
discretize this space. We are able to discretize this space, again
as a consequence of the choice of $\Omega$. The exponential
component nature of $\Omega$ provides for momentum kicks of
exactly one atomic unit of momentum in the direction of
propagation of the laser light (the $x$-direction). Thus the only
momentum kicks that are not in a single unit of atomic momentum in
the $x$-direction are due to spontaneous emission of photons from
the atom. This allows for a momentum kick of one atomic unit in
any direction, which, when projected onto the $x$-direction,
allows for a random kick of anywhere between $-1$ and $1$ atomic
unit. However the relative infrequency of the spontaneous emission
allows us to approximate this by a kick of $-1$, $0$, or $1$.

This approximation requires us to convert the integral over the
atomic dipole distribution to a discrete sum. Following Ref.
\cite{Hen01a}, we let
\begin{equation} \int^1_{-1}W(u)du \longrightarrow
\sum_{u=-1,0,1}V(u),
\end{equation} where the discrete function $V(u)$ is obtained by
making sure that it has the same normalization, mean momentum kick
and mean squared momentum kick. This gives $V(-1)=\frac{1}{5}$,
$V(0)=\frac{3}{5}$ and $V(1)=\frac{1}{5}$.

We now need to define our matrix notation. The simulations here
will set $\hbar$, $k$ and $m$ to 1 so that the kinetic energy
operator can be represented as \begin{equation}
\frac{p_x^2}{2m}\longrightarrow \frac{1}{2}
\sum_{n=-25}^{25}n^2\ket{n}\bra{n},\end{equation} where $\ket{n}$
denotes a momentum eigenstate. The $e^{-ikx}$ operator in a
momentum Hilbert space just acts as a raising operator given by
\begin{equation} R=\sum^{25}_{n=-25}\ket{n+1}\bra{n}.
\end{equation} The $e^{ikx}$ operator is just the lowering
operator $R^\dagger$. For simplicity, we take the initial density
matrix to be the zero momentum state $\ket{0}$, given by a
$51\times51$ matrix of zeros with a 1 in the very centre.

These are all the approximations required to allow the simulation
of the adiabatic approximations. The full master equation and the
secular and dressed-state approximations require the internal
state information as well. This is achieved by constructing the
tensor product of the $51\times51$ momentum Hilbert with a
$2\times2$ internal state Hilbert space. This is easily done using
the quantum optics toolbox.

The only problem remaining is to adopt a method of simulation.
There are a number of different simulation methods available as
part of the quantum optics toolbox as well as any number of
methods available using regular ordinary differential equation
techniques. The method we use here is a built-in function of the
Quantum Optics Toolbox called {\tt odesolve}. This allows options
to be specified for use on both smooth and stiff ODE problems. The
simulations presented here using {\tt odesolve} were checked using
first, a hybrid Euler and matrix exponential method, and second, a
modified mid-point method combined with Richardson extrapolation
from Press {\it et al} \cite{Press}. The results all agreed well
but the {\tt odesolve} method was by far the fastest and easiest
to use.

We wish to simulate the experimental regime of Hensinger {\it et
al}, where
$\delta\gg\Omega_\mathrm{max}\gg\Gamma\sim\Omega_\mathrm{max}^2
/\delta$. However, the actual experimental parameters would be
prohibitively time-consuming to simulate. This is both because of
the separation in time scales between the fastest ($\delta$) and
slowest dynamics ($\Gamma\Omega_{\rm max}$), and also because of
the basis size required. The latter is determined by the fact that
 $p^2_{\mathrm{max}}/2m$
must be larger than the effective potential drop,
$\Omega^2_{\mathrm{max}}/4\delta$. If we were to use the
parameters of the experiment of \cite{Hen01a} then we would need a
basis size of more than $\pm70\hbar k$.
However, we can scale the parameters down and still
 preserve the regime of the
experiments.

As well as working in the
regime
$\delta\gg\Omega_\mathrm{max}\gg\Gamma\sim\Omega_\mathrm{max}^2
/\delta$, we require for the validity of the adiabatic
approximations that $\Gamma$
be much larger than the oscillation frequency near the
potential minimum, $\omega_\mathrm{osc}$.
In our scaled units ($\hbar=k=m=1$),
the latter is
of order $\sqrt{\Omega_\mathrm{max}^2/\delta}$. On this basis,
we have chosen paramaters of $\delta=10^4$,
$\Omega_\mathrm{max}=2\times10^3$ and $\Gamma=200$, leaving
$\Omega_\mathrm{max}^2/\delta=400$, giving $\omega_\mathrm{osc}=20$.
For Rb as in Ref.~\cite{Hen01b}, we have $k=2\pi/780$nm
and $m=1.42\times10^{
-25}$kg, so in SI units, the frequency unit is $\hbar k^{2}/m =
4.821\times10^4$ s$^{-1}$. Note that the $\Gamma$ we have chosen is
{\em not} the true radiative decay rate for Rb.

The scaled time unit can be given meaning by examining the
spontaneous emission rate. For each of the approximations, the
fluorescence rate is (to leading order) $\Gamma\overline{\Omega^2}/4
\delta^2$. Here $\overline{\Omega^{2}}$
is a time-averaged effective value, which would be somewhat less than
$\Omega_{\rm max}^{2}$.
This means that after a time period of $4\delta^2
/\Gamma \Omega_{\rm max}^2$, we would expect there to have somewhat
under one spontaneous emission. This time is 2 scaled time units. There is,
however, a lot of evolution occurring in that time period. To
compare the approximations, we look in detail at a period from 0
to 2 time units, and also look at the long time results at 8 time
units.

\section{Results of the simulations\label{Results}}

In comparing the results of the simulations, we first compare the
accuracy of the approximations, and then the resources required to
perform the calculations.

\subsection{Accuracy\label{Accuracy}}

To compare the accuracy of the simulations, we look at the
momentum distribution of the atom as it evolves through time. The
interesting components of this evolution are the probability to
have 0 momentum and the probability to have 1 atomic unit of
momentum as the atom evolves through time. The other probabilities
evolve similarly to one of these two.

\begin{figure}\includegraphics[width= .4\textwidth]{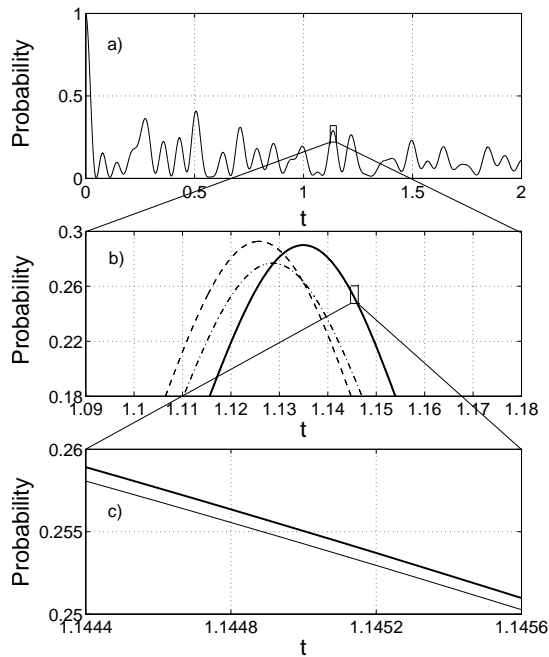}
\caption{\label{full0}The probability to have zero momentum versus
scaled time. a) Overall picture of the probability to have zero
momentum. Individual lines are not resolvable at full size. b) The
heavy solid line is the full master equation, the dashed line
represents the standard adiabatic approximation while the dot-dash
line represents the secular approximation. The full master
equation is indistinguishable from the more sophisticated
adiabatic and dressed state approximations even at this size. c)
The thin solid line represents the dressed-state approximation and
the sophisticated adiabatic approximation.}
\end{figure}

Firstly, we will look at the probability to have zero momentum
over a relatively short timescale. The approximations are so close
to the full master equation that, at full size, they are almost
impossible to distinguish from the full master equation. Fig.
\ref{full0}a shows an overall picture of how the probability to
have zero momentum evolves through
time.

Fig. \ref{full0}b zooms in on a section of the full size figure to
illustrate the differences between the approximations. As we can
see, the secular and standard adiabatic approximation evolutions
both significantly lead that of the full master equation in time.
The dressed-state and sophisticated adiabatic approximations are
very close to the full master equation solution even at this
magnification. Fig \ref{full0}c zooms in even closer to try to
distinguish the sophisticated adiabatic approximation from the
full master equation.

\begin{figure}
\includegraphics[width= .4\textwidth]{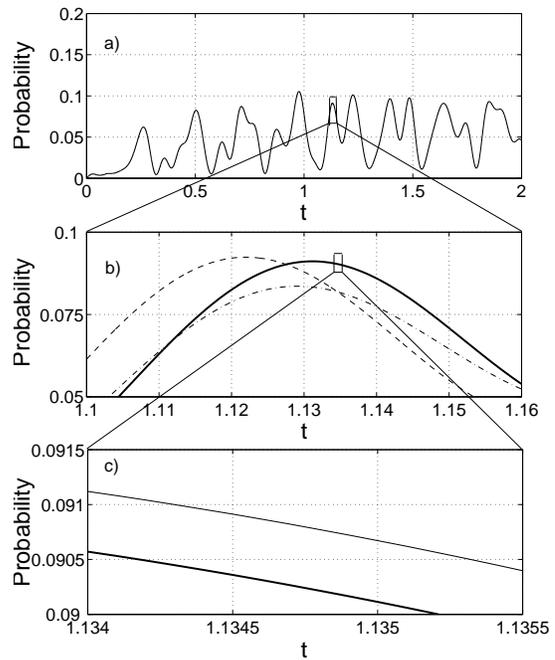}
\caption{\label{full1}The probability to have 1 atomic unit of
momentum versus scaled time. a) Overall picture of the probability
to have 1 atomic unit of momentum. Individual lines are again not
resolvable at full size. b) The heavy solid line is the full
master equation, the dashed line represents the standard adiabatic
approximation while the dot-dash line represents the secular
approximation. The full master equation is indistinguishable from
the more sophisticated adiabatic and dressed state approximations
even at this size. c) The thin solid line represents the
dressed-state approximation and the sophisticated adiabatic
approximation.}\end{figure}

This trend continues as we investigate the probability to have 1
atomic unit of momentum in Fig. \ref{full1}. The inaccuracy of the secular and
standard adiabatic approximations are again evident in Fig.
\ref{full1}b. The more sophisticated adiabatic approximation is
very close to the full master equation evolution and is still
indistinguishable at this magnification.

Fig. \ref{full1}c provides a means of comparing the sophisticated
adiabatic approximation to the full master equation evolution in
detail. As we can see, the more sophisticated adiabatic
approximation is again very close to the full master equation.

\begin{figure}\includegraphics[width=
.4\textwidth]{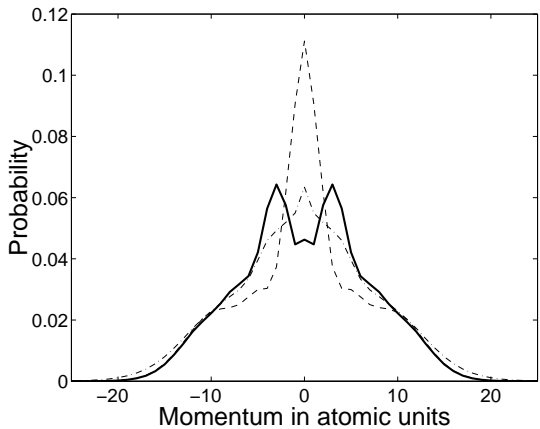}\caption{\label{long}Long time probability
density. The solid line represents the full master equation which
is indistinguishable from the dressed-state and more sophisticated
adiabatic approximations at this magnification. The dashed line
represents the standard adiabatic approximation while the dash-dot
line is the evolution described by the secular approximation. It
is worth noting here that due to our approximations, the
probability density is only defined for integer values of
momentum. We have \textquotedblleft connected the
dots\textquotedblright to aid the eye.}
\end{figure}

The last visual comparison to make is to see how the evolution
described by the approximations matches that of the full master
equation after a very long time period. The time period that has
elapsed in Fig. \ref{long} is 8 time units, after which, we would
have expected there to be a number of spontaneous emissions. At
this point in time, we compare the probability densities described
by the approximations and the full master equation.
 It is evident from Fig. \ref{long} that the standard
adiabatic and the secular approximations are quite poor methods
for simulating the full master equation evolution over a long time
period. The main reason for this is the leading behaviour such as
we see in Fig. \ref{full0}a. Even at long times, the probability
to have zero momentum is still oscillating and the leading
behaviour means that the standard adiabatic and secular
approximations are not oscillating in phase with the full master
equation. Thus, even though they follow roughly the same shape,
they are not necessarily at the same point in the oscillation as
the full master equation. What is even more striking though is
that the secular approximation evolution seems to follow a
slightly different trend for the probabilities to have larger
momentum. The intriguing features of the secular approximation are
discussed in Sec \ref{Discussion}.

\begin{figure}\includegraphics[width=
.4\textwidth]{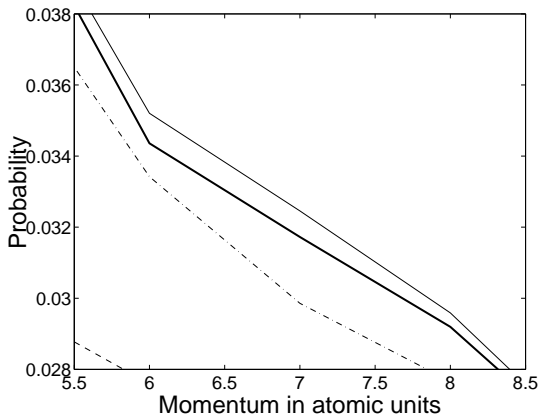}\caption{\label{longzoom}Long time
probability density. The heavy solid line represents the full
master equation evolution while the thin solid line represents the
sophisticated adiabatic and dressed-state approximations. Again
the secular and standard adiabatic approximations are represented
by the dot dash line and the dashed line
respectively.}\end{figure}

To examine how close the sophisticated adiabatic and the
dressed-state approximations are to the full master equation,
Fig.~\ref{longzoom} focuses on a smaller section to provide a
comparison.  It is evident from Fig.~\ref{longzoom}
that the dressed-state and sophisticated adiabatic approximations
are very close to the full master equation, even at this long
time.

\subsection{Resources\label{Resources}}

The main resource required to perform the simulations is time.
Although each approximation has different memory and processing
power requirements, these needs are reasonably accurately
reflected in the time each simulation takes to run. The times
quoted in table \ref{table} are the times required to calculate
the set of results from 0 to 8 time units and are quoted in
seconds.
\begin{table}[h]\caption{\label{table}Comparison of the
times each simulation requires to
run.}\begin{ruledtabular}\begin{tabular}{lc} Simulation & Time (s) \\
\hline Full Master equation  & 617.7 \\
Standard Adiabatic Approximation & 16.98 \\ More Sophisticated
Adiabatic Approximation & 18.31 \\ Secular Approximation & 76.43 \\
Dressed State Approximation\footnote{The time quoted here is idential for the
more sophisticated adiabatic approximation because the equations are
identical for an atom initially in the ground state.} & 18.31
\end{tabular}\end{ruledtabular}
\end{table}

As we can see from the results in table \ref{table}, the adiabatic
approximations are both clearly the fastest. The secular
approximation is just over 4 times larger than the adiabatic
approximations. This is not entirely unexpected. We would have
expected at least a doubling in time by using the methods that
included state information. The secular approximation is supposed
to force the coherences to zero. Unfortunately, using our method
to simulate this allows zero to be anywhere up to $10^{-5}$. This,
although small, still has to be processed explaining the four-fold
increase in time. One reason it is just over 4 times the time
required for the adiabatic approximations could be that after the
evolution has been calculated, there is still a partial trace to
be performed to obtain a solution of the same form as the
adiabatic approximations. We could have limited this time by
simulating two coupled equations instead of a full master equation
and then would probably have only doubled the time taken.

\section{\label{Discussion}Discussion}

The standard and more sophisticated adiabatic approximations take
similar times to simulate, and are much faster than the full
master equation. The only difference between them is an extra
potential term. It turns out, though, that this Hamiltonian term
is quite important in accurately describing the motion of the
atom. While the standard approach evolves too quickly and leads
that of the full master equation evolution, the more sophisticated
approach with the modified potential does not suffer this problem.
The evolution described by this more sophisticated approach is
very close to the full master equation, even at long times. Of
course the dressed-state approximation offers the same accuracy as
the sophisticated adiabatic approximation which may be useful if
we wanted to simulate an initially excited atom.

These successes contrast the results from the secular
approximation. As one can see from
Figs.~\ref{full0}~and~\ref{full1}, the secular approximation not
only leads the full master equation solution but it also predicts
a lower probability to have either zero or one atomic unit of
momentum. This result is surprising because the secular
approximation master equation is quite similar to the others.
Investigating this further, we find that the secular approximation
master equation simulation shows that the probability to have 25
atomic units of momentum increases exponentially much faster than
any of the other approximations. To analyse this in another
manner, we find that for the secular approximation Tr$[\rho]$
falls off from 1 exponentially much faster than any of the other
approximations or the full master equation.

If we consider the dressed-state approximation master equation,
\erf{finaldressME}, we notice that the Hamiltonian terms are, to
leading order, the same as the final secular approximation master
equation \erf{secularME}. The only advantage the dressed state
approximation has in terms of it's Hamiltonian is the presence of
the higher order term. The leading order Lindbladian terms are
also identical with only the higher order terms differing. The
dressed-state master equation includes Lindbladian terms which
give rise to a momentum kick to the atom (from the $\mathcal{B}$
superoperator) without a change in the internal state of the atom.
The higher order secular approximation master equation Lindbladian
term provides the opposite. Here there is an internal state change
without a momentum kick to the atom. Thus it is quite unexpected
that the secular approximation solution should differ so greatly
from the dressed-state approximation solution as they are
identical in the leading order terms.

It is a fairly simple matter to investigate the effect of the
differences between the two approximations by simulating only
parts of the master equations. If we only included the Hamiltonian
terms from each master equation, the difference would only be the
higher order term in the dressed state master equation. This term
has the same effect here as it does for the sophisticated
adiabatic approximation, correcting the leading behaviour. It does
not however account for the secular approximation predicting lower
probabilities as it does in Figs.~\ref{full0} and \ref{full1}.
Thus to investigate the difference between the higher order
Lindbladian terms, we drop the $\mathcal{B}$ superoperator from
the dressed-state master equation and add it to the secular
approximation master equation. This does not correct the problems
with the secular approximation nor does it severely affect the
dressed-state approximation. This only leaves the inherent
difference that the dressed-state master equation provides a
correction term without requiring a state change where the secular
approximation forces a state change in it's correction term. Thus
we have to conclude that the best approximation to the full master
equation involves a Lindbladian term that does not change the
internal state of the atom. Lacking it, the secular approximation
is the poorest.

\begin{figure}
\includegraphics[width=.4\textwidth]{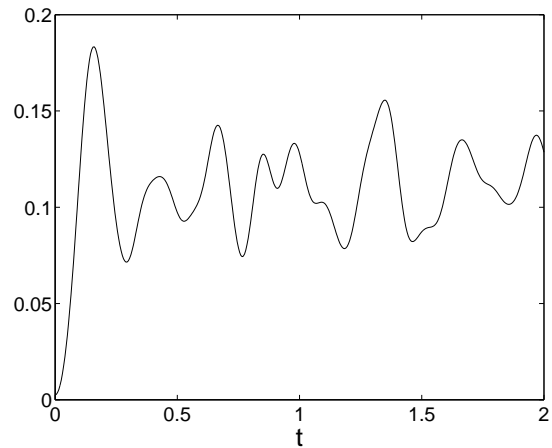}
\caption{\label{fig5} Ratio of
Tr$[\rho_{ee}{p_x^2}/{2m}]$ to Tr$[\Gamma\rho_{ee}]$ for
the sophisticated adiabatic approximation. All other simulations
look similar.}\end{figure}

One other point to note in this discussion is one of the
conditions on which the adiabatic elimination techniques are
based. This is that $\Gamma\rho_{ee}$ be much larger than
$\mathcal{K}\rho_{ee}$. This will be satisfied if
\begin{equation}
\mathrm{Tr}\left[ \rho_{ee} \frac{p_x^2}{2m}\right] /
\Gamma\mathrm{Tr}\left[\rho_{ee}\right]\ll 1.
\label{checkpsquared}\end{equation} For the parameter regime of
this investigation, it is not obvious that this holds. Fig
\ref{fig5} shows how the ratio in \erf{checkpsquared} evolves
through time for the sophisticated adiabatic approximation.
Here we see that numerically, this
ratio is around 0.1 which is of the same order as the other ratios
(such as $\Omega_{\rm max}/\delta$) which are required to be
small for our approximations.

Finally, we discuss the possibility of simulating with the true
experimental parameters. This is difficult because of the stiffness of
the full master equation, and the basis size required for all methods.
The latter problem can be avoided by using
quantum trajectories \cite{MolCasDal93, DumZolRit92,Car93}.
Hensinger {\it et al} \cite{Hen01a} actually use quantum
trajectory simulations based on the secular approximation master
equation. It is however, possible to convert any master equation
of the Lindblad form to a quantum trajectory simulation
\cite{Wiseman}. All of the approximate master equations we have
developed here have been written in the Lindblad form and as such
all of these could be simulated using quantum trajectories.

\section{conclusion}

There are a number of theoretical models for
 the motion of an atom as it interacts with a light field.
This paper has investigated the possibility of using four
different approximations as opposed to using the full master equation to
simulate an experimental system. Two have been widely used in the past and
two have not. We have given a detailed
explanation of the mathematical principles to perform each of these
approximations on a fairly general system. We have also compared them
numerically to   to the true dynamics from the full master equation.

In a regime of particular experimental interest, we have found that
the most accurate results are obtained from two approaches that we have
introduced here, a sophisticated adiabatic approach and a dressed-state
approach. These give identical equations in the regime of interest, and
in terms of resources, they are almost as fast to simulate as
 the standard
adiabatic approximation. This has been most used in the past,
but deviates significantly from the true dynamics for long times.
The other approximation that has been used in the past, the secular
approximation, is even poorer. On top of the failings of the
standard adiabatic approximation, it takes longer to simulate and
appears to  produce anomalous momentum
diffusion.

\acknowledgements{We gratefully acknowledge discussions with W.
Hensinger which prompted us to pursue the investigation. This work
was supported by the Australian Research Council.}

\end{document}